\documentclass[aps, prd, twocolumn, superscriptaddress, nofootinbib]{revtex4}
\usepackage{graphicx}
\usepackage{dcolumn}
\usepackage{amssymb,amsmath,amsthm,mathrsfs}

\begin{document}


\newcommand{\Eq}[1]{Eq. \ref{eqn:#1}}
\newcommand{\Fig}[1]{Fig. \ref{fig:#1}}
\newcommand{\Sec}[1]{Sec. \ref{sec:#1}}

\newcommand{\PHI}{\phi}
\newcommand{\vect}[1]{\mathbf{#1}}
\newcommand{\Del}{\nabla}
\newcommand{\unit}[1]{\mathrm{#1}}
\newcommand{\x}{\vect{x}}
\newcommand{\ScS}{\scriptstyle}
\newcommand{\ScScS}{\scriptscriptstyle}
\newcommand{\xplus}[1]{\vect{x}\!\ScScS{+}\!\ScS\vect{#1}}
\newcommand{\xminus}[1]{\vect{x}\!\ScScS{-}\!\ScS\vect{#1}}
\newcommand{\diff}{\mathrm{d}}

\def\Dia{\Diamond}\def\lab{\label}\def\lan{\langle}\def\lar{\leftarrow}
\def\lf{\left}\def\lrar{\leftrightarrow}
\def\Lrar{\Leftrightarrow}\def\noi{\noindent}
\def\non{\nonumber}\def\ot{\otimes}\def\pa{\partial}\def\ran{\rangle}
\def\rar{\rightarrow}\def\Rar{\Rightarrow}
\def\ri{\right}\def\ti{\tilde}\def\we{\wedge}\def\wti{\widetilde}
\def\al{\alpha}\def\bt{\beta}\def\ga{\gamma}\def\Ga{\Gamma}
\def\de{\delta}\def\De{\Delta}\def\ep{\epsilon}
\def\ze{\zeta}\def\te{\theta}\def\Te{\Theta}\def\ka{\kappa}
\def\la{\lambda}\def\La{\Lambda}\def\si{\sigma}\def\Si{\Sigma}
\def\om{\omega}\def\Om{\Omega}
\def\AB{{_{A,B}}}\newcommand{\mlab}[1]{\label{#1}}
\def\CP{{_{C\!P}}}\def\T{{_{T}}}
\def\AB{{_{A,B}}}\def\mass{{_{1,2}}}
\def\flav{{e,\mu}}\def\1{{_{1}}}\def\2{{_{2}}}
\def\bp{{\bf {p}}}\def\bk{{\bf {k}}}\def\br{{\bf {r}}}\def\bx{{\bf {x}}}\def\bQ{{\bf {Q}}}
\def\bl{{\bf {l}}}\def\bq{{\bf {q}}}\def\bj{{\bf {j}}}
\def \ak{\alpha^r_{{\bf k},e}(0)}\def \akd{\alpha^{r\dag}_{{\bf k},e}(0)}
\def\ap{\alpha^s_{{\bf p},e}(0)}\def\apd{\alpha^{s\dag}_{{\bf p},e}(0)}
\def\br{{\bf{r}}}\def\bI{{\bf{I}}}
\def\fourint{\int\!\!\!\int\!\!\!\int\!\!\!\int}\def\threeint{\int\!\!\!\int\!\!\!\int}
\def\twoint{\int\!\!\!\int}
\def\disp{\footnotesize disp}
\def\Qo{{_{Q\!,1}}}\def\Qt{{_{Q\!,2}}}\def\ko{{_{k\!,1}}}\def\kt{{_{k\!,2}}}
\newcommand{\ide}{1\hspace{-1mm}{\rm I}}

\newcommand{\be}{\begin{equation}}
\newcommand{\ee}{\end{equation}}
\newcommand{\bea}{\begin{eqnarray}}
\newcommand{\eea}{\end{eqnarray}}
\newcommand{\vu}{{\mathbf u}}
\newcommand{\ve}{{\mathbf e}}
\newcommand{\vk}{{\mathbf k}}
\newcommand{\vx}{{\mathbf x}}
\newcommand{\vy}{{\mathbf y}}

\newcommand{\uden}{\underset{\widetilde{}}}
\newcommand{\den}{\overset{\widetilde{}}}
\newcommand{\denep}{\underset{\widetilde{}}{\epsilon}}

\newcommand{\nn}{\nonumber \\}
\newcommand{\dd}{\diff}
\newcommand{\fr}{\frac}
\newcommand{\del}{\partial}
\newcommand{\eps}{\epsilon}
\newcommand\CS{\mathcal{C}}

\def\be{\begin{equation}}
\def\ee{\end{equation}}
\def\ben{\begin{equation*}}
\def\een{\end{equation*}}
\def\bea{\begin{eqnarray}}
\def\eea{\end{eqnarray}}
\def\bal{\begin{align}}
\def\eal{\end{align}}


\title{Neutrino oscillations and superluminal propagation, in OPERA
or otherwise}

\newcommand{\addressImperial}{Theoretical Physics, Blackett Laboratory, Imperial College, London, SW7 2BZ, United Kingdom}

\author{Jo\~{a}o Magueijo}
\affiliation{\addressImperial}


\date{\today}

\begin{abstract}
We digress on the implications of recent claims of superluminal neutrino
propagation. No matter how we turn it around such behaviour 
is very odd and sits uncomfortably even within ``far-fetched'' theories. 
In the context of non-linear realizations of the Lorentz group 
(where superluminal misbehaviour is run of the mill) one has to accept
rather contrived constructions to predict superluminal  
properties for the neutrino.
The simplest explanation is to require that at least 
one of the mass states be 
tachyonic. We show that due to neutrino mixing, 
the flavor energy does not suffer from the usual runaway pathologies
of tachyons. For non-tachyonic mass states the theories become more 
speculative. A neutrino specific dispersion relation is exhibited,
rendering the amplitude of the effect reasonable for a standard Planck
energy. This uses the fact that the beam energy is close to the geometrical
average of the neutrino and Planck mass; or, seen in another way, 
the beam energy is unexceptional but its gamma factor is very large.
A dispersion relation crossing over from a low energy bradyonic branch
to a high energy tachyonic one is also considered. We comment on 
consistency with SN 1987A within these models.
\end{abstract}

 \pacs{To be done}

\maketitle


No particle can claim to have caused as much havoc in modern
physics as the neutrino. Ever since its proposal by Pauli in 1930,
the neutrino has been the source of relentless mischief, with its
parity violating properties supplying the prime example. 
The propensity of neutrino flavors 
to oscillate comes a close second. Neutrino oscillations have now 
evolved from the wild idea first suggested by Pontecorvo~\cite{Pont} 
into a respectable fact. Against this historical background
one should perhaps not be 
overly surprised by recent claims that superluminal speeds have been achieved 
by neutrinos~\cite{opera}. As with any experimental result, 
these claims should be taken with a macroscopic grain of salt. 
The potential for uncharted systematic errors can never be 
overemphasized~\cite{percneut,contaldi}. Nonetheless,
theorists are easily excitable creatures who don't require much 
prompting before going down the ``what if'' alley~\cite{giov,caccia}. 
The purpose of this paper is to follow this time honored tradition, 
exploring the implications of the possible superluminal nature of the
neutrino for the general theory of oscillations, and also for theories
proposing faster than light propagation in cosmology~\cite{vsl}. 

It is undeniable that the alleged neutrino superluminal behaviour 
sits uncomfortably in our theoretical constructions, even those tailored to
accommodate the strangest behaviour. Faster than light propagation has been
envisaged in the context of varying speed of light theories, in particular
those where the speed of light is energy dependent~\cite{dsr,dsr1}. 
The latter may be 
realized in a variety of ways, for example via a deformation of the
usual dispersion relations, then made invariant by adopting
a non-linear representation of the Lorentz 
group~\cite{dsr2}. This approach was actually applied to 
neutrinos a few years back~\cite{bmp1,bmp2}, where it was shown that 
{\it with minimal assumptions}, and {\it without assuming tachyonic behaviour
for the mass states}, it is possible for the flavor states to display
features peculiar to tachyons, e.g. regarding the
end point of beta decay (see, e.g.~\cite{katrin}). 
However, it was stressed that in such theories
{\it the neutrino would always remain slower than light}. In this paper
we identify the assumptions that must be broken for superluminal behaviour 
to be unleashed. We start with the construction of~\cite{bmp1,bmp2}, 
but then consider more general theories as well. A measure of contrived
success is achieved.

The quantum field theory of neutrinos has been the subject of some debate.
We do not want to wed our considerations to this debate, but since one
of our models results from this discussion we briefly review the issues
at stake. It was shown in~\cite{BV95} that it is possible to define a 
Hilbert space of flavor states, with a well defined vacuum state and creation 
and  annihilation operators. Comparison with the usual treatment shows
that this approach amounts to a different choice of vacuum. The matter
is far from pedantic and corrections to the usual formula for neutrino
oscillations arise~\cite{BHV99}. Flavor states constructed in this way are 
eigenstates of the flavor charge and the momentum operators but are
not eigenstates of the Hamiltonian~\cite{BHV99,bosonmix}. Nonetheless
one can compute the expectation value of the Hamiltonian $H$ on the 
flavor states and define from it dispersion relations. These dispersion
relations are not the usual hyperbolic ones and fall within the remit of
deformed special relativity, as explained in~\cite{bmp1}. 
One may then wonder under what circumstances these
theories could accommodate superluminal neutrino speeds, in line
with the theories considered in~\cite{dsr,dsr1,dsr2}.

To illustrate our point we shall consider the simplifying case of 
two flavors, and examine Pontecorvo states
(thereby neglecting a number of subtleties with the set up
of the flavor Hilbert space; see~\cite{bmp2}). We then 
have the generic mixing relations for flavors $a$ and $b$ 
(which could be the muon and the tauon or the electron and the muon):
\bea\non |\nu_a\ran& = &\cos\theta \,\,|\nu_1\ran +
\sin\theta\,\, |\nu_2\ran
\\ [2mm] \label{Pontec}
|\nu_b\ran & =& -
\sin\theta\,\, |\nu_1\ran + \cos\theta\,\,|\nu_2\ran \; .  \eea
Without loss of generality we take $0\le\te\le\frac{\pi}{4}$ and
$m_2 > m_1$. Computing the expectation value of the Hamiltonian on the
flavor states (\ref{Pontec}) we find:
\bea E_a &\equiv & \langle\nu_a| H|\nu_a \rangle  = \om_{k,1}
\cos^2\te + \om_{k,2} \sin^2 \te
\label{dispersion}\\ [2mm] 
E_b &\equiv& \langle\nu_b| H|\nu_b \rangle  =  \om_{k,2}
\cos^2\te + \om_{k,1} \sin^2 \te\label{dispersion2}
\eea
where $H|\nu_i\rangle=\omega_i|\nu_i\rangle$ ($i=1,2$) and \be
\omega_{k,i} = \sqrt{ {\mathbf k}^2+m_i^2}\,  .\ee Since
the sum of two square roots is generally not a square root,
except for trivial cases, flavor states do not satisfy 
hyperbolic dispersion relations. If we take seriously the flavor states
thus constructed, and if we wish to avoid a conflict with the principle
of relativity, we should therefore postulate that flavor states transform
according to a non-linear representation of the 
Lorentz group~\cite{dsr1,dsr2}. Following the derivation in~\cite{bmp1}, 
the dispersion relations~(\ref{dispersion}) may be cast in the form:
\bea \label{lorentz} E_a^2 \, f_a^2(E_a)\, - \,\bk^2
\,g_a^2(E_a)\, =\,  M_a^2
\\ [2mm] \label{lorentz2}
E_b^2 \, f_b^2(E_b)\, -\, \bk^2\, g_b^2(E_b)\, =\,
M_b^2\; .
\eea
from which the recipes given in~\cite{dsr2} for the construction of
the non-linear representation are 
straightforward to apply. Specifically, Eq.~(\ref{dispersion}) leads to
\bea
\label{eqe1}
 \left(E_a^2 - \om_1^2 \cos^4\te - \om_2^2 \sin^4 \te\right)^2 =  4 \om_1^2
 \om_2^2 \sin^4 \te \cos^4\te
 \\ [1.5mm]
\left(E_b^2 - \om_2^2 \cos^4\te - \om_1^2 \sin^4 \te\right)^2 =
4 \om_1^2
 \om_2^2 \sin^4 \te \cos^4\te
 \eea
from which one finds
\bea 2f^2_a(E_a) &=&1+ \frac{1} {\cos^2(2\te) }
\nonumber
\\
&&- \frac { \sqrt{E_a^2 + \left(m_2^2 - m_1^2\right) \cos(2\te)
}} {E_a} \tan^2(2\te)
\\
g^2_a(E_a)& =& 1
\\
M_a^2&=&\frac {{\tilde m}_-^2}{\cos 2\te}=
\frac{m_1^2\cos^4\te -m_2^2\sin^4\te} {\cos (2\te)}\; .
\eea
The above assumes $\theta\ne\pi/4$.
For maximal mixing, $\theta=\pi/4$, we have instead:
\bea f^2_a(E_a) &=&1+ \left(\frac{m_1^2-m_2^2} {4E_a^2}\right)^2\\
g^2_a(E_a)& =& 1\\
M_a^2&=&\frac {m_1^2+m_2^2} {2}
 \eea
Similar expressions, but with $m_1 \lrar m_2$, apply to flavor $b$.
A number of corrections arise from a more careful treatment~\cite{bmp2}, 
but they do not qualitatively change the rest of our discussion.

It was noted in~\cite{bmp1} that even with non-tachyonic mass
states ($m_i^2>0$) it is possible to obtain $M_a^2< 0$ for
\be \label{tachcond}
 \tan\theta> \sqrt{m_1/m_2}\; .
\ee
This has a number of interesting implications concerning the end point
of beta decay, which can indeed be as one would expect from a tachyonic
neutrino. But the interesting point made in~\cite{bmp1} is that this
can be accomplished without the usual pathologies of tachyons. Unlike
tachyons, there is not a minimal non-vanishing momentum, for which 
the energy is zero, connecting the positive and negative branch of the 
dispersion relations thereby
leading to a runaway instability. On the contrary it is easy to check that
the flavor states can have $k=0$, at which point they reach their minimal 
non-zero energy:
\bea E_a^{min} &=&
m_1 \cos^2\te + m_2 \sin^2 \te\label{emin1a}
\\ 
E_b^{min} &=& 
m_2 \cos^2\te + m_1 \sin^2 \te \; .\label{emin1b}
\eea
In addition these dispersion relations do not entail superluminal
propagation. They imply a propagation velocity (assuming 
the concept applies to flavor states), which is a monotonically growing 
function of the momentum which always satisfies $v<1$. This is 
true for both phase speed $v=p/E$ or  group speed $v=dE/dp$, 
and is hardly news. Starting from (\ref{dispersion}) and (\ref{dispersion2})
we may rewrite the dispersion relations in 
the ultra-relativistic regime as:
\bea
E_a\approx k+\frac{{\tilde m}_a^2}{2k}&,{\quad}&
{\tilde m}_a^2=m_1^2\cos^2\theta+m_2^2\sin^2\theta\label{mass1}\\
E_b\approx k+\frac{{\tilde m}_b^2}{2k}&,{\quad}&
{\tilde m}_a^2= m_2^2\cos^2\theta+m_1^2\sin^2\theta\; .\label{mass2}
\eea
Non-tachyonic mass states therefore ensure non-tachyonic behavior for the 
flavor states in the relativistic regime. 
Regrettably (or perhaps not) the tachyons found in~\cite{bmp1}
only behave like tachyons at low energies, and even then within a limited
scope.  The implication is that some assumption
in this very minimal model must be dropped if we want to accommodate 
superluminal speeds.  
This result is not very surprising. With the very conventional assumptions
we have made one may take the view that the only states with a well 
defined (phase or group) speed are the mass eigenstates. If none of these
are tachyonic then no superluminal behavior can be expected. 

A possible new ingredient is to postulate that one
of the mass eigenstates is tachyonic. Then superluminal behavior can be
expected, even if no other assumptions are changed. The formalism we have
presented, however, introduces an interesting novelty. We find
that neutrino oscillations cure
a fundamental tachyonic pathology. Continuing to illustrate our point with
the simplified model presented above, let us assume that the first mass
state is tachyonic, i.e. $m_1^2=-a^2<0$. 
Neutrino mixing superposes particles with the same momentum, so 
we are forced to conclude
that {\it both}
flavor states, $a$ and $b$, have a non-vanishing minimum momentum,
$k_{min}=a$, just like any tachyon. However, unlike standard tachyons, 
the energy at this point is not zero, but the minimal energy:
\bea E_a^{min} &=&
E_a(k=a)=\sqrt{a^2+m_2^2} \sin^2 \te\
\\ 
E_b^{min} &=& E_b( k =a)=
\sqrt {a^2+m_2^2} \cos^2\te \; . 
\eea
As with standard particles, or ``bradyons'',
there is of course a negative energy branch, but
there is a gap between the two preventing
runaway instabilities. Given that interactions are mediated via
flavor states, we can argue that it is their dispersion relations, 
and not those of the mass eigenstates, which are relevant for stability
discussions. Therefore an unsavory property of tachyons seems to have
been removed from the problem. This is an interesting twist on 
the tachyonic neutrinos of~\cite{glash}.

In spite of this novelty, the flavor states are not regular bradyons.
As one can read off from (\ref{mass1}), if:
\be
\tan\theta<\frac{a}{m_2}
\ee
the flavor $a$ has tachyonic properties at {\it high} energies, unlike the case
discussed in~\cite{bmp1}. Assuming the tachyonic behaviour seen by~\cite{opera}
is due to a mass 
eigenstate we should have $a\sim 0.1$Gev, which is uncomfortably
high. With a flavor state taking the blame the condition would 
become more flexible:
\be
{\tilde m}_a^2=m_1^2\cos^2\theta+m_2^2\sin^2\theta\sim 
-(0.1{\rm Gev})^2\; .
\ee
Choosing a single tachyonic mass state, may lead to other problems, 
such as a detrimental effect on the coherence length of the neutrino 
(but note that oscillation formulas have 
to be rederived in this case). However, the argument
above can be repeated if more than one mass eigenstate is tachyonic.
In the two flavor example we have chosen, if $m_1^2=-a^2<0$ 
and $m_2^2=-b^2<0$, with $b>a$, then the minimal momentum is 
$k_{min}=b$, for which the energies are 
\bea E_a^{min} &=&
E_a(k=b)=\sqrt{b^2-a^2} \cos^2 \te\
\\ 
E_b^{min} &=& E_b( k =b)=
\sqrt {b^2-a^2} \sin^2\te \; ,
\eea
and the argument still carries through.

Is there any way of achieving the same effect with non-tachyonic mass
states within the usual theory of oscillations? We were unable to find
one, but what we have said so far suggests obvious modifications to the 
theory, capable of accomplishing the task. Suppose that at some energy
$E_0$ the flavor states become eingenstates of the Hamilton endowed with
the dispersion relations (\ref{lorentz})-(\ref{lorentz2}). 
Suppose further that at this scale $f_a(E_a)$
goes to 1 faster than $1+\frac{C}{k^2}$, say by replacing
\be
f_a\rightarrow 1+(f_a-1)e^{-E/E_0}\; .
\ee
Then, at low energies everything 
we have said still stands, but at high energies we fail to witness the
transition from mass $M_a^2<0$ (assuming condition
(\ref{tachcond})) to mass ${\tilde m}_a^2>0$. 
Quite the opposite: the mass remains $M_a^2<0$ at high 
energies. Therefore, as in~\cite{bmp1}, the minimal energy of
flavor $a$ is $E_a^{min}$ given by
(\ref{emin1a}), and is achieved with $k=0$. But then we must cross-over 
from the bradyonic to the tachyonic quadrant. As the 
energy increases, the neutrino goes from $v<1$ to break the speed of
light barrier at $E\sim E_0$, reaching a maximal speed before 
slowing down to $v=1$ from above, as the energy goes to infinity. 
As in the cases discussed above, no instabilities are present, but
obviously superluminal behavior is obtained.
Once we have accepted this sort of dispersion relation we
could postulate something similar for the mass eigenstates and keep
the standard theory of oscillations. Note that with the assumption that
the flavor states diagonalize the eigenstates at high energies ($E>E_0$)
we would predict no oscillations. 

Once we go beyond conventional theory, why not tie in these results
with other non-conventional theories. However, 
as was pointed out in~\cite{giov}, appealing to Planck scale 
effects known to raise the speed of light does 
not blend nicely with the observations
of~\cite{opera}. Suppose, as an example, that we take
\be
E^2- \frac{p^2}{\left(1-\frac{p}{E_P}\right)^2} =m^2
\ee
for the base dispersion relations, applicable to all particles, and
so presumably also to the neutrino mass eigenstates. 
This theory has $E_P$ as the maximal 
momentum and displays a varying speed of light~\cite{dsr2}. 
This propagates into the speed for a neutrino mass state $m$ as:
\be
v\approx 1+\frac{2E}{E_P} -\frac{m^2}{2E^2}\; .
\ee
For a standard neutrino scenario the last term is negligible
($\sim 10^{-19}$) and so, looking at the first term, 
we'd need the Planck mass to be of order $10^6$Gev (rather than the
conventional $10^{19}$Gev) to explain the observations.
Even then we'd need to explain why the effect hadn't been observed 
in other particles, well explored in this energy range. The neutrino
would have to be unique in its probe of this dispersion relation.

While we agree with~\cite{giov} in this respect\footnote{As shown 
in that reference, changing
the powers in the dispersion relations may soften this problem, but never 
repeals it.}, we note that once 
we accept that the neutrino would have to have unique dispersion
relations, not shared by other particles, the situation improves.
Indeed then it would be possible to explain
the magnitude of the observed 
effect without making the Planck scale unduly small.
This could be done with a sort of ``seesaw mechanism''. 
As an illustrative example, suppose that neutrino mass states
feel dispersion relations of the type:
\be
\frac{E^2-k^2}{1+\alpha^2\frac{E^4}{m^3E_P}}=m^2\; ,
\ee
where $\alpha$ is a dimensionless parameter.
Since the dispersion relation is specific to neutrinos we shouldn't
be alarmed to see $m$ appear in the deformation, as well as $E_P$.
Then superluminal effects would
kick in for $E\sim\alpha\sqrt{m M_P}$. Moreover, we 
could accommodate the results
in~\cite{opera} with $\alpha\sim 1$ (i.e. no fine tuning), 
since the neutrino speed would satisfy:
\be
\frac{v-c}{c}=-\frac{m^2}{2E^2}+3\alpha^2\frac{E^2}{mM_P}\; .
\ee
Again the first term is negligible in the context of~\cite{opera}
while the second explains the observation.
The crucial aspect in this argument is that the energy
in the neutrino beam ($\sim 10$GeV) misses the geometrical average 
of the neutrino mass scale (say $0.1$eV) and the standard Planck scale
($\sim 10^{19}$GeV) by a factor of order $\sim 10^{-5}$: precisely 
the observed fractional
superluminal propagation speed. We may not want
to indulge in numerology, but we can make use of it.

In summary we have presented a number of arguments,
from the point of view of varying speed of light theories, which could lead
to neutrino superluminal behavior. 
In all honesty none of these are very palatable. 
Appealing to at least a single tachyon eigenstate might be the 
simplest way out. The highlight of this paper was the discovery that
in such a set up, due to neutrino mixing, the flavor eigenstates
need not suffer from the instabilities peculiar to tachyons, 
provided they are interpreted as non-linear representations of the 
Lorentz group and masses and mixing angles are chosen carefully. 

Beyond that our paper delved into more speculative ideas 
(by now, however, the neutrino is expected to be crazy; 
see e.g.~\cite{paes1,cpt,nikos,pfeifer}).
We may force the flavor dispersion relations we have found well 
beyond existing theory. In the most extreme case this would
entail a cross-over from the bradyonic branch at low energies to the
tachyonic branch at high energies. Then, the neutrino (either a flavor or 
a mass state, both options are up for grabs) starts off 
subluminal at low energies, but then, as its energy increases, 
``breaks the speed
of light'' to reach a maximal speed at a given finite
energy. Its speed then decreases with energy, 
approaching the speed of light
from above, as the energy goes to infinity, like any other good tachyon.
As long as this is interpreted within 
the framework of non-linear representations
of the Lorentz group, causality violations are not necessarily
implied, since one
should employ the associated non-linear Lorentz 
transformations~\cite{rainb,dagny}.
It is a far-fetched idea, but extreme circumstances call for extreme measures. 

We also showed how invoking Planck scale physics (and its habit
of speeding up light) would imply a ridiculously 
low Planck energy scale, as well as the embarrassing question: why the 
neutrino and not other particles? We suggested a possible solution to the first
problem by simply accepting the second. With a 
set of dispersion relations specifically tailored for the neutrino, 
we were able to prove that the observed effect could be predicted with
a standard Planck scale. Central to the argument is the fact that whilst the
energy of the neutrino beam is unexceptional, its gamma factor is very large.
By putting the neutrino mass into the expression for the deformed dispersion
relations we can then put this fact to good use and 
predict a velocity $10^{-5}$ above the speed of
light for the meagre $E\sim 10$GeV, whilst keeping the Planck scale
$E_P\sim 10 ^{19}$GeV.

We could have gone further, elaborating on the important point made 
in the last paragraph. We note that until recently, 
Ultra-High-Energy cosmic rays held the record for the largest
gamma factor:
\be
\gamma_{\rm UHECR}\sim 10^{11}\; .
\ee
Even for a neutrino mass state with $m\sim 0.01$eV this would have been 
achieved in OPERA, with the lowest mass states definitely improve on the 
mark. With MINOS~\cite{minos} 
and OPERA we are finally probing:
\be
\gamma_\nu\gg 10^{11}.
\ee 
Therefore,  
even though the energy scales in OPERA are unexceptional, something is
unique.  
Had the neutrino been well behaved, {\it this would have been the closest we 
ever got to the speed of light, without actually sitting on it}. Instead 
the neutrino
broke the light barrier. 
In a separate
paper, we propose a theory building on this fact. The basic idea
is that one may extend the non-linear representations of the Lorentz group
developed in~\cite{dsr1,dsr2} with a construction where the 
angle $\xi$ in Lorentz transformations is replaced by a new
${\tilde \xi}=\xi_0\tanh(\xi)$, saturating at a given gamma factor. 
We stress that,  since the Lorentz group is non-compact, 
its experimental testing is necessarily open-ended.

An obvious concern (for theorists and experimentalists alike) 
is consistency with other observations, namely 
SN1987A (e.g.~\cite{ellis,caccia}).
Our theoretical considerations are very much at the level of toy models,
so we cannot offer a detailed examination. Nonetheless we close by presenting
some thoughts.
Face value our first model (a tachyonic mass state) contradicts SN1987A. 
If one adopts the view that the mass states, rather than the flavor states, 
are what ``moves'' independently in vacuum, and postulates that 
one of them is tachyonic, then there should have been a tachyonic signal 
from SN1987A arriving 3-4 years earlier. However, the 
matter is far from clear cut. Firstly, 
how sure are we that such a ``premonitory''
burst didn't arrive? Then, assuming the experiments were up and 
running~\cite{fargion}, 
could we arrange for a scenario where we could reasonably have missed 
this early burst?
The answer hinges on a crucial detail. The SN1987A are mainly 
electron neutrinos whereas those at OPERA are muonic neutrinos. Suppose 
the mixing of the former with the tachyonic state is negligible, unlike 
the latter. Then some tachyonic signal from SN1987A would have 
arrived, but it could have been negligible. Even though there are muon
and tauon neutrinos among the thermal neutrinos, these are a minority,
and the signal arriving 3 or 4 years ahead could easily have 
passed unnoticed.

The same argument can be adapted to
our third model. Deformed dispersion relations
should in principle apply to mass states (although the idea may be combined
with a stronger ``individuality'' for flavors). One could therefore design a
theory (either with very different masses, or with different
$\alpha$ for the different mass states) where the proposed mechanism for 
speeding up neutrinos is much stronger for one mass state than the others. 
Requiring a
low mixing with electron neutrinos would then suppress the faster than
light signal from SN1987A. Some thermal muon and tauon neutrinos
are produced (and most of the observed neutrinos come from this pool),
but the statistics aren't good enough to rule out a tachyonic signal
from them.  At any rate, we stress that the argument on thermal neutrinos 
results from computer simulations, not observations, and these 
would have to
be revised, for consistency, should deformed dispersion relation
affect at least one state.

As for our second model, it actually {\it predicts} that tachyonic 
behaviour should not be seen in the SN1987A. All that needs to be
done is for the cross-over from bradyonic to tachyonic branch to
happen above 10MeV. This is similar in spirit to one of the proposals
in~\cite{caccia}.

The most conservative explanation of OPERA, of course, remains to dismiss the
the results as an ``experimental error'', the ultimate 
bail out of the theorist . At the end of the day
it is up to experimentalists to sort out their
wares. We wait for an independent experiment with bated breath.

{\bf Acknowledgements}
I'd like to thank J. Halliwell, L. Lyons, 
J. Sedgbeer and  D. Wark for discussions.


\end{document}